\documentclass[conference]{IEEEtran}
\IEEEoverridecommandlockouts
\usepackage{cite}
\usepackage{amsmath,amssymb,amsfonts}
\usepackage{algorithmic}
\usepackage{graphicx}
\usepackage{textcomp}
\usepackage{comment}
\usepackage{xcolor}
\usepackage{todonotes}
\usepackage{enumitem}
\usepackage{booktabs}
\usepackage{multirow}
\usepackage{xspace}
\usepackage{standalone}
\usepackage{amsmath}
\usepackage{hyperref}
\usepackage{tikz}
\usetikzlibrary{positioning, arrows.meta, decorations.pathmorphing}

\def\BibTeX{{\rm B\kern-.05em{\sc i\kern-.025em b}\kern-.08em
    T\kern-.1667em\lower.7ex\hbox{E}\kern-.125emX}}

\newcommand{\sepsis}{\emph{\textbf{Sepsis}}\xspace}
\newcommand{\bpia}{\emph{\textbf{Bpic2012\_A}}\xspace}
\newcommand{\bpib}{\emph{\textbf{Bpic2012\_B}}\xspace}
\newcommand{\fines}{\emph{\textbf{Traffic\_Fines}}\xspace}
\newcommand{\lstm}{\emph{\textbf{lstm}}\xspace}
\newcommand{\lstmi}{\emph{\textbf{lstm1}}\xspace}
\newcommand{\lstmii}{\emph{\textbf{lstm2}}\xspace}
\newcommand{\baseline}{\emph{\textbf{train\_log}}\xspace}
\newcommand{\cvae}{\emph{\textbf{cvae}}\xspace}
\newcommand{\mb}[1]{\boldsymbol{\mathrm{#1}}}

\begin{document}

\title{
Generating the Traces You Need: A Conditional Generative Model for Process Mining Data\\
\thanks{We acknowledge the support of the PNRR project FAIR - Future AI Research (PE00000013), under the NRRP MUR program funded by the NextGenerationEU, the HORIZON 2020 HumanE-AI project (Grant 952026), and the PRIN project PINPOINT Prot. 2020FNEB27, CUP H23C22000280006 and H45E2100021000. 
}
}

\author{    \IEEEauthorblockN{Riccardo Graziosi\textsuperscript{\textsection}}
		\IEEEauthorblockA{\textit{Fondazione Bruno Kessler, Trento, Italy} \\
			rgraziosi@fbk.eu}
   \and
            \IEEEauthorblockN{Massimiliano Ronzani\textsuperscript{\textsection}}
            \IEEEauthorblockA{\textit{Fondazione Bruno Kessler, Trento, Italy} \\
			mronzani@fbk.eu}
		\and
   \IEEEauthorblockN{Andrei Buliga}
		\IEEEauthorblockA{\textit{Fondazione Bruno Kessler, Trento, Italy} \\
			\textit{Free University of Bozen-Bolzano, Bolzano, Italy}\\
			abuliga@fbk.eu}
   		\and
		\IEEEauthorblockN{Chiara Di Francescomarino}
		\IEEEauthorblockA{\textit{University of Trento, Trento, Italy} \\
			c.difrancescomarino@unitn.it}
		\and
   		\IEEEauthorblockN{Francesco Folino}
		\IEEEauthorblockA{\textit{ICAR-CNR, Rende, Italy} \\
			francesco.folino@icar.cnr.it}
		\and
		\IEEEauthorblockN{Chiara Ghidini}
		\IEEEauthorblockA{\textit{Free University of Bozen-Bolzano, Bolzano, Italy} \\
			chiara.ghidini@unibz.it}
		\and
            \IEEEauthorblockN{Francesca Meneghello}
		\IEEEauthorblockA{\textit{Fondazione Bruno Kessler, Trento, Italy} \\
			\textit{Sapienza University of Rome, Rome, Italy}\\
			fmeneghello@fbk.eu}
   \and
   		\IEEEauthorblockN{Luigi Pontieri}
		\IEEEauthorblockA{\textit{ICAR-CNR, Rende, Italy} \\
			luigi.pontieri@icar.cnr.it}
}

\maketitle
\begingroup\renewcommand\thefootnote{\textsection}
\footnotetext{Equal contribution}
\endgroup
\begin{abstract}
In recent years, trace generation has emerged as a significant challenge within the Process Mining community.
Deep Learning (DL) models have demonstrated accuracy in reproducing the features of the selected processes.
However, current DL generative models are limited in their ability to adapt the learned distributions 
to generate data samples based on specific conditions or attributes.
This limitation is particularly significant because the ability to control the type of generated data can be beneficial in various contexts,
enabling a focus on specific behaviours, exploration of infrequent patterns, or simulation of alternative ``what-if'' scenarios.

In this work, we address this challenge by introducing a conditional model for process data generation based on a conditional variational autoencoder (CVAE). 
Conditional models offer control over the generation process by tuning input conditional variables, enabling more targeted and controlled data generation. 
Unlike other domains, CVAE for process mining faces specific challenges due to the multiperspective nature of the data and the need to adhere to control-flow rules while ensuring data variability.
Specifically, we focus on generating process executions conditioned on control flow and temporal features of the trace, allowing us to produce traces for specific, identified sub-processes.
The generated traces are then evaluated using common metrics for generative model assessment, along with additional metrics to evaluate the quality
of the conditional generation.

\end{abstract}

\begin{IEEEkeywords}
Process Mining, Deep Learning, Generative AI, Conditional models
\end{IEEEkeywords}

\section{Introduction}
\label{sec:Introduction}

Process mining (PM) \cite{DBLP:books/sp/Aalst16} is 
a research field that focuses on the analysis, monitoring, and improvement of business processes based on event logs.
Within this field, generative models have emerged in recent years as crucial tools for generating new event trace samples that replicate process behavior\cite{DBLP:conf/caise/TaxVRD17, EVERMANN2017129, MM_Pred, Evermann_Fettke, Tax2018AnIC, BINet, Camargo_lstm, LSTM_GAN}. 
These models support a range of applications, including
anomaly detection \cite{BINet, vae_anomaly_detection}, predictive monitoring \cite{DBLP:conf/caise/TaxVRD17}, what-if scenario analysis \cite{CAMARGO2020113284} and conformance checking \cite{Sani_2020}.
An important yet underexplored aspect of trace generation is the ability to produce traces that follow different distributions from the training data, allowing exploration of various dimensions of interest within the process.
These dimensions may include exploring 
\emph{what-if} scenarios, expanding variants of interest (especially when significant for the process analysis but numerically low), or exploring 
resource contingency plans.


According 
\cite{CamargoDumas_survey}, generative models can be categorized into two main families: Data-Driven Process Simulation (DDPS) and Deep Learning (DL).
DDPS constructs explicit process models from data, esnuring that complete information about the simulation is always available.
These models are beneficial for providing insights into specific subprocesses
and allow to modify almost every aspect of the simulation. 
However, DDPS often relies on oversimplified assumptions, leading to unrealistic simulations and 
data generation. Moreover, they struggle to capture long-term dependencies.
DL models are statistical models that accurately capture the correlations between 
features in the generated samples. Despite their accuracy, DL models are ``black box'' systems, making it challenging to transparently expose the underlying process model. More importantly, existing DL-based generative models are rigid, limiting the generation of distributions different from the training data. This constraint significantly inhibits 
the exploration of specific scenarios or dimensions of interest.
Hybrid models, which integrate the accuracy of DL techniques with the transparency of explicit process models, have been recently introduced \cite{dsim, rims}. While they may have potential to generate specific data for dimensions of interest, an assessment of this capability is still missing.


Conditional generative models \cite{NIPS2015_8d55a249} have been proven effective in various domains to mitigate the rigidity observed in DL generative models.
These models generate outputs influenced by certain input variables. This offers a means to guide the generation process based on desiderata for the expected output.

In this work, we introduce a conditional variational autoencoder (CVAE) model for generating traces based on LSTM neural networks.
Compared to other domains, 
developing a CVAE for process mining presents two main challenges:
\begin{itemize}
\item Process execution data have an inherently multi-perspective nature. Traces consist of sequence of temporal features with both categorical (events) and numerical (timestamps) characteristics. This complexity is further increased by the inclusion of payloads.
Moreover, all these components are strongly interconnected. 
Therefore, due to their diverse nature, each component of the 
generated samples requires a dedicated module in the model architecture,
which must still produce a coherent output.
\item While it is desirable for the generative model to produce data with some variability, event sequences must adhere to causal constraints and diverse contextual factors. Thus, the variability must remain consistent with process constraints to ensure meaningful results.
\end{itemize}

A further contribution of our work is the proposal of 
a novel evaluation methodology for assessing process generation quality in a conditional context.
Alongside common metrics for evaluating generative process models, such as the accuracy of the control flow and generated timestamps \cite{Dumas_framework},
we introduce a new analysis to measure the impact of the conditioning variable and measure the actual conditioning rate.
Moreover, unlike many simulation scenarios that aim to closely replicate 
historical data, generative models in a conditional context should introduce variability in the generated output 
while remaining within process constraints. Thus, our evaluation framework includes metrics to assess the variability and conformance of the generated traces with the process rules.


By integrating 
conditional models into process data generation, this work aims to enhance the flexibility and control of DL generative models in Process Mining. Evaluation on four different examples based on three real-world event logs shows promising results. 
Specifically, the generated traces are accurate, exhibit good variability, comply with process constraints, and demonstrate that the conditional generative model effectively controls the types of traces produced.


\section{Background}
\label{sec:Background}
In this section we introduce the main concepts useful to understand the remainder of the paper.

\subsection{Event Log}
An event log $\mathcal{L}$ records the executions of a business process in terms of execution \emph{traces}.
A trace $x$ 
consists of a sequence of ordered events $x = \langle e_{1}, e_{2}, \ldots e_{n} \rangle$. Events are characterized by multiple attributes (\emph{data attributes}): primarily, an event refers to an activity label and a timestamp indicating when the activity was executed.
It may also include information about the resource executing or initiating the activity, and other data attributes. 
Some attributes are static and consistent throughout the trace's execution, 
known as \emph{trace attributes}.
Data associated with events and traces in event logs are also called \emph{data payloads}.
We can represent a trace $x$ as:
\begin{equation}\label{eq:trace}
\begin{aligned}
x = \langle (a_1, T_1, \mathbf{d}_1), \dots, (a_n, T_n, \mathbf{d}_n)\rangle
\end{aligned}
\end{equation}
where $a_i$ is the $i$-th activity executed in the trace, $T_i$ is its timestamp, and $\mathbf{d}_i$ is a vector containing its data payload, including static trace attributes.

\subsection{Conditional Variational Autoencoders (CVAEs)}
Autoencoders are neural network architectures used for unsupervised learning tasks~\cite{doi:10.1126/science.1127647}. They consist of an encoder, \(E_\phi\), and a decoder, \(D_\theta\), representing non-linear transformations parametric to $\theta$ and $\phi$, respectively. 
The encoder maps the input data \(x\) into a latent representation \(\mb{z}=E_\phi(x)\), while the decoder output \(\hat{x}=D_\theta(\mb{z})\) should reconstruct the original input from the latent representation. 
Mathematically, an autoencoder aims to minimize the reconstruction error between the input and the output:
\begin{equation}
\mathcal{L}_{\text{AE}}(x, \theta, \phi) = J_{rec}(x,\hat{x}).
\label{eq:ae_loss}
\end{equation}
\noindent where $J_{rec}$ denotes some kind of distance/error function over the data space (e.g., $\lVert x-\hat{x} \rVert^2$ in the case  $x,\hat{x} \in \mathbb{R}^N$).

Variational autoencoders (VAEs)~\cite{kingma2022autoencoding} lift this basic autoencoder architecture to the level of 
a generative model where $x$ and $\mb{z}$ are interpreted as observed and latent random variables, respectively, such that the joint distribution 
of $x$ and $\mb{z}$ is factored as $p_\theta(x | \mb{z}) \cdot p(\mb{z})$, where $p_\theta(x | \mb{z})$ is a distribution to be learned, and the latent prior $p(\mb{z})$ is typically set to a standard multivariate Gaussian distribution (i.e., $p(\mb{z}) = \mathcal{N}(\Vec{0}, \mb{I})$, where $\mb{I}$ is an identity matrix).\footnote{This allows for easily generating any $\hat{x}$ using a two-phase scheme: first $\mb{z}$ is sampled from the latent prior $p(\mb{z})$, and then $\hat{x}$ is sampled from $p(x \mid \mb{z})$.}

VAEs are trained to minimize the (expectation over the real data distribution of the) following negative \emph{Evidence Lower Bound} (ELBO) $\mathcal{L}_{\text{VAE}}$, consisting of a reconstruction loss term (echoing that in Eq. \eqref{eq:ae_loss}) plus a regularization term, which is computed as the Kullback-Leibler (KL) divergence between a learned latent distribution 
\(q_\phi(\mb{z}|x)\) and the latent prior \(p(\mb{z})\), with a factor $\beta$ controlling the strength
of the regularization:
\begin{equation}
\mathcal{L}_{\text{VAE}}(x, \theta, \phi) = J_{\text{VAE}}(x,\theta,\phi) + \beta \cdot \text{KL}\big(q_\phi(\mb{z} | x) \mid \mid p(\mb{z})\big)
\label{eq:vae_loss}
\end{equation}
where \(\phi\) and \(\theta\) denote the parameters of the encoder and decoder sub-nets, now modelling the learned distributions \(q_\phi(\mb{z} | x)\) and \(p_\theta(x | \mb{z}) \), respectively, while the reconstruction term is $J_{\text{VAE}}(x,\theta,\phi) = - \mathbb{E}_{\mb{z} \sim q_\phi(\mb{z} | x)} \ln p_\theta(x | \mb{z})$.

Conditional variational autoencoders (CVAEs) incorporate conditional information into both the encoder and the decoder. In CVAEs, both the encoder and the decoder take the input data \(x\) and conditioning variables \(c\) as inputs~\cite{NIPS2015_8d55a249}.
The conditional variable 
$c$ represents the specific condition or attribute that guides the generation process, enabling the model to produce data samples with desired characteristics.
The encoder maps $x$ and $c$ to the parameters of a (variational) posterior distribution $q_\phi(\mb{z}\vert x,c)$ of the latent variables \(\mb{z}\),
while the decoder models a distribution $p_\theta(x | \mb{z},c)$ for generating/reconstructing the input data conditioned on both the latent representation and the conditioning variables. 
Optimal parameters for both the encoder and decoder are learned by  
minimizing (the expectation over the real data of) the following negative ELBO objective:
\begin{equation}
\mathcal{L}_{\text{CVAE}}(x, c, \theta, \phi) = J_{\text{CVAE}} + \beta \cdot \text{KL}(q_\phi(\mb{z} \mid x,c) \mid\mid p(\mb{z}\mid c))
\label{eq:cvae_loss}
\end{equation}
\noindent where \(q_\phi(\mb{z} | x,c)\) is the (learned) conditional posterior distribution of the latent variables (given both the observed data and conditioning variables), 
$J_{\text{CVAE}}(x,c,\theta,\phi) = - \mathbb{E}_{\mb{z} \sim q_\phi(\mb{z} | x,c)} \ln p_\theta(x | \mb{z},c)$, 
and $p(\mb{z}|c)$ is a prior distribution (conditioned on $c$) for the latent variables $\mb{z}$. 
Usually, for any given $x$ and $c$, $q_\phi(\mb{z} | x,c)$ is assumed to be a multivariate Gaussian distribution with a diagonal covariance matrix. 



\section{Related work}
\label{sec:Related_work}

Generative DL models have been extensively studied in recent years in the PM field.
The primary idea behind most of these models, in the context of 
predictive process monitoring, is to generate a trace by iteratively predicting the next activity for a prefix.
For instance, \cite{DBLP:conf/caise/TaxVRD17} introduces an LSTM-based method \cite{LSTM_paper} that generates the remaining sequence of events with associated timestamps, given a trace prefix. This approach uses one-hot encoding to represent activities, which struggles with high-dimensional inputs, i.e., processes with a large number of activities.

In \cite{EVERMANN2017129}, an LSTM neural network is used to generate event sequences, addressing the dimensionality issues with embeddings for activities. However, this method does not handle numerical features and thus cannot generate event timestamps.
Other approaches for activity sequence generation include an LSTM-based method in \cite{MM_Pred}, n-grams encoding with neural networks in \cite{Evermann_Fettke}, and Markov models, RNN, and automata-based models in \cite{Tax2018AnIC}. Despite their variety, these methods share a common limitation: they do not generate timestamps.

In \cite{BINet} and \cite{vae_anomaly_detection}, generative methods based on GRU neural networks and variational auto-encoders respectively, have been applied for anomaly detection.

In \cite{Camargo_lstm}, the authors combine elements from prior work to build an accurate LSTM-based generative model that generates events with timestamps and associated roles, and can produce traces from scratch using an ``hallucination'' mechanism.
To ensure sufficient variability in the generated traces, the selection of the next events is performed using a random sampling method based on the predicted probability distribution outputted by the model.
While this method increases the variability of the generated traces, it may occasionally produce traces  inconsistent with the global process distribution.
In \cite{LSTM_GAN}, 
an LSTM model for predicting the next event and its timestamp, is trained adopting a generative adversarial network (GAN) approach.

A comparison in \cite{CamargoDumas_survey} between these two methods and a variant of \cite{Camargo_lstm} with GRU layers shows that the original LSTM-based method in \cite{Camargo_lstm} achieves the best results on average.
\footnote{In fact, as discussed in \cite{gans_2024,gans_mc_2023}, while GANs excel at generating realistic samples (e.g., high-fidelity photos, deep fakes), they often focus on limited portions of the data distribution and are prone to the notorious mode collapse problem, for which a general, 
effective, solution is still missing.
By contrast, VAEs are effective in modeling multimodal distributions, which may well occurr in process logs, while providing control in the generative process~\cite{gans_2024}. 
}

In this work we provide a further contribution to the state-of-the-art in trace generation (events, timestamps and trace attributes) by employing a conditional variational auto-encoder (CVAE) based on LSTM neural networks.
This approach has two main advantages:
(i) it allows generating traces with good variability by sampling from the latent space, without adopting weighted random choices on the LSTM output;
(ii) it adds control to the generation process by setting the conditional variables.
To our knowledge, this is the first DL conditional model applied to trace generation in PM.

\section{Approach}
\label{sec:Approach}
In this section we present a conditional variational autoencoder (CVAE) architecture for the generation of traces. In this work, we consider the generation of traces with only activities $a_i$, timestamps $T_i$, and static traces attributes, which can be numerical $\mathbf{d}_{num}$ and categorical $\mathbf{d}_{cat}$. The trace \eqref{eq:trace} can then be rewritten as
$x=\{(\mathbf{d}_{num}, \mathbf{d}_{cat}) , \langle (a_1, T_1), \dots, (a_n, T_n)\rangle\}$. 

\begin{figure*}[!t]
    \centering
\resizebox{0.98\linewidth}{!}{\begin{tikzpicture}[
  node distance=1.5cm and 2cm,
  box/.style={draw, fill=gray!20, minimum width=2cm, minimum height=1cm, align=center, text=black, font=\Large},
  textnode/.style={align=center, font=\Large},
  every node/.style={font=\Large} 
]


    \node[draw=gray, rounded corners=8pt, minimum width=20cm, minimum height=8cm, align=center, line width=0.5mm, anchor=north west] at (-1.8cm,2cm) (box_lhs) {};
    \node[below] at ([yshift=-3mm, xshift=20mm] box_lhs.north west) {\textbf{ENCODER}};
    \node[draw=gray, rounded corners=8pt, minimum width=16.3cm, minimum height=8cm, align=center, line width=0.5mm, anchor=north west] at ([xshift=5cm]box_lhs.north east) (box_rhs) {};
    \node[below] at ([yshift=-3mm, xshift=20mm] box_rhs.north west) {\textbf{DECODER}};

\node[textnode] (ai) {$a_i$};
\node[textnode, below=1cm of ai] (ti) {$t_i$};
\node[textnode, below=1cm of ti] (cat_attr) {$\mathbf{d}_{cat}$};
\node[textnode, below=1cm of cat_attr] (num_attr) {$(T_1, \mathbf{d}_{num})$};

\node[box, right=1.5cm of ai] (embed1) {Embed};
\node[box, below=of embed1, yshift=-7mm] (embed2) {Embed};

\draw[->, >=Triangle] (ai) -- (embed1);
\draw[->, >=Triangle] (cat_attr) -- (embed2);

\node[box, right=1.5cm of embed1, yshift=-5mm] (lstm1) {LSTM};
\node[box, right=1.5cm of embed2, yshift=-5mm] (linear1) {Linear};

\draw[->, >=Triangle] (embed1) -- (lstm1);
\draw[->, >=Triangle] (embed2) -- (linear1);

\draw[->, >=Triangle] (ti) to (lstm1);
\draw[->, >=Triangle] (num_attr.east) to (linear1);

\node[textnode, right=1.5cm of lstm1, yshift=-1.5cm, node distance=3cm] (concatenated) {concatenate};
\node[textnode, below=2.3cm of concatenated] (label) {conditional\\variable $c$};

\draw[->, >=Triangle] (lstm1.east) -- (concatenated.north west);
\draw[->, >=Triangle] (linear1.east) -- (concatenated.south west);
\draw[->, >=Triangle] (label.north) -- (concatenated.south);

\node[box, right=1.5cm of concatenated, yshift=1.5cm] (linear2a) {Linear};
\node[box, below=2.3cm of linear2a] (linear2b) {Linear};

\draw[->, >=Triangle] (concatenated.north east) -- (linear2a.west);
\draw[->, >=Triangle] (concatenated.south east) -- (linear2b.west);

\node[textnode, right=1.5cm of linear2a] (mu) {$\mu$};
\node[textnode, right=1.5cm of linear2b] (sigma) {$\sigma$};

\draw[->, >=Triangle] (linear2a) -- (mu);
\draw[->, >=Triangle] (linear2b) -- (sigma);

\node[textnode, right=1.5cm of mu, yshift=-1.5cm] (n_mu_sigma) {Sampling:\\$z\sim\mathcal{N}(\mu, \text{diag}(\sigma))$};
\node[textnode, below=1.9cm of n_mu_sigma] (label2) {conditional\\variable $c$};

\draw[-, dashed, >=Triangle] (mu) -- (n_mu_sigma.west);
\draw[-, dashed, >=Triangle] (sigma) -- (n_mu_sigma.west);
\draw[-, dashed, >=Triangle] (label) -- (label2);

\node[box, right=1.5cm of n_mu_sigma, node distance=3cm] (linear3) {Linear};

\draw[->, >=Triangle] (n_mu_sigma) -- (linear3.west);
\draw[->, >=Triangle] (label2.north east) -- (linear3.south west);

\node[textnode, right=1.5cm of linear3] (zu) {$z_U$};

\draw[->, >=Triangle] (linear3) -- (zu);

\node[box, right=1.5cm of zu, yshift=3cm] (lstm2) {LSTM};
\node[box, right=1.5cm of zu, yshift=1cm] (lstm3) {LSTM};
\node[box, right=1.5cm of zu, yshift=-1cm] (linear4) {Linear};
\node[box, right=1.5cm of zu, yshift=-3cm] (linear5) {Linear};

\draw[->, >=Triangle] (zu) -- (lstm2.west);
\draw[->, >=Triangle] (zu) -- (lstm3.west);
\draw[->, >=Triangle] (zu) -- (linear4.west);
\draw[->, >=Triangle] (zu) -- (linear5.west);

\draw[->, >=Triangle] (lstm2) -- (lstm3);

\node[box, right=1.5cm of lstm2] (linear6a) {Linear};
\node[box, right=1.5cm of lstm3] (linear6b) {Linear};
\node[box, right=1.5cm of linear4] (linear6c) {Linear};
\node[box, right=1.5cm of linear5] (linear6d) {Linear};

\draw[->, >=Triangle] (lstm2) -- (linear6a);
\draw[->, >=Triangle] (lstm3) -- (linear6b);
\draw[->, >=Triangle] (linear4) -- (linear6c);
\draw[->, >=Triangle] (linear5) -- (linear6d);

\node[textnode, right=1.5cm of linear6a] (hat_ai) {$\hat{a}_i$};
\node[textnode, right=1.5cm of linear6b] (hat_ti) {$\hat{t}_i$};
\node[textnode, right=1.5cm of linear6c] (hat_cat_attr) {$\hat{\mathbf{d}}_{cat}$};
\node[textnode, right=1.5cm of linear6d] (hat_num_attr) {$(\hat{T}_1, \hat{\mathbf{d}}_{cat})$};

\draw[->, >=Triangle] (linear6a) -- (hat_ai);
\draw[->, >=Triangle] (linear6b) -- (hat_ti);
\draw[->, >=Triangle] (linear6c) -- (hat_cat_attr);
\draw[->, >=Triangle] (linear6d) -- (hat_num_attr);

\end{tikzpicture}}
    \caption{A high level overview of the CVAE for process mining. Activities $a_i$ and event interarrival times $t_i=T_i-T_{i-1}$ are incrementally processed by the LSTM subnet, while all the categorical $\mathbf{d}_{cat}$ and numerical $\mathbf{d}_{num}$ attributes of $x$ (including trace arrival time $T_1$) are processed by the Linear subnet.}
    \label{fig:cvae}
\end{figure*}
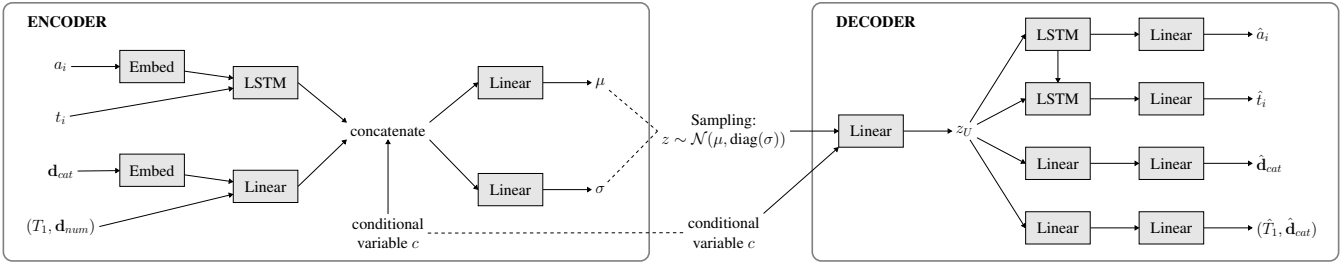

In order to be able to apply
the CVAE to business processes, we had to adapt it by employing encoder and decoder architectures suitable for handling both sequential data (control flow and timestamps) and non-sequential data (trace attributes). The model is depicted in Figure \ref{fig:cvae}. 


In particular, for sequential data, we drew inspiration from seq2seq models, which are commonly used in NLP, where an encoder model maps a variable-length input sequence to a fixed size vector, which is then ``unrolled'' back to a variable-length sequence by a decoder model \cite{sutskever2014sequence}.

The following sections explain the preprocessing steps, the encoder and decoder architectures, the training approach, and the generation process of new traces.

\subsection{Preprocessing}
\label{sec:approach-preprocessing}
In the preprocessing phase a trace $x$ is properly manipulated to be feed to the encoder.
Each event's timestamp $T_i$ is decomposed into two parts: (i) the \emph{trace arrival time} $T_1$, which corresponds to the timestamp of the first event in the trace, and is handled by the model in the same way as a numerical trace attribute; and (ii) the \emph{event interarrival time} $t_i := T_i - T_{i-1}$. The \emph{event interarrival time} is normalized using the 95th percentile value instead of the maximum value. While this causes some values to be outside the $[0, 1]$ range, we found it useful in practice in order to ignore outlier timestamps, i.e. very high interarrival time, that could cause the normalization to squash \emph{event interarrival times} to an exceedingly narrow range close to zero. 
Numerical trace attributes are preprocessed by normalizing them to the $[0, 1]$ interval using min-max normalization. 

\subsection{Encoder}
\label{sec:approach-encoder}
The goal of the encoder is to map any trace $x$ and conditioning variable $c$ to the mean vector $\mb{\mu}_{x,c}$ and variance vector $\mb{\sigma}_{x,c}$ that fully specify the multivariate Gaussian distribution $q_\phi(\mb{z} | x,c)\equiv \mathcal{N}(\mb{\mu}_{x,c},\text{diag}(\mb{\sigma}_{x,c}))$, with a diagonal covariance matrix, for the latent variables $\mb{z}$ (the subscripts in $\mb{\mu}_{x,c}$ and $\mb{\sigma}_{x,c}$ will be omitted whenever the dependency of these parameters on $x$ and $c$ is clear from the context).
Firstly, in the encoder each categorical variable (activities and categorical attributes) is passed through its own embedding layer and transformed into a numerical vector.
Then, the encoder consists of two paths: the first makes use of an LSTM layer to handle activities and timestamps, the second employs a fully connected layer for each trace attribute. Thereafter, outputs of the two paths are concatened together, the conditional variable is added, and lastly two fully connected layers are used to map to mean $\mb{\mu}$ and variance $\mb{\sigma}$ vectors.

\subsection{Decoder}
\label{sec:approach-decoder}
After sampling a latent space vector $\mb{z}$ from the Gaussian distribution $\mathcal{N}(\mb{\mu}_{x,c},\text{diag}(\mb{\sigma}_{x,c}))$ identified by the encoder, the goal of the decoder is to generate a trace from $\mb{z}$ that is as similar as possible (modulo a certain level of variability) to the original trace $x$. 

First of all, since we are in a conditional setting, the conditional variable $c$ is concatenated to $\mb{z}$.
Then, a fully connected layer upsamples $\mb{z}$ to a higher dimension vector $\mb{z}_{U}$. This approach, even though it has been criticized in previous NLP works \cite{bowman2015generating}, has proven effective in our case to improve decoder performances.
Activities and timestamps are obtained from $\mb{z}_U$ using two different autoregressive LSTMs. 
At each time step $i$, the activity LSTM takes as input both $\mb{z}_U$ and the previously reconstructed activity $\hat{a}_{i-1}$ (for the first time step, a special End Of Trace (EOT) token is used) and outputs the current activity $\hat{a}_i$. 
The timestamp LSTM takes as input, in addition to $\mb{z}_U$ and the previous event interarrival time $\hat{t}_{i-1}$, also the current reconstructed activity $\hat{a}_i$, and outputs the current event interarrival time $\hat{t}_i$.

Different configurations has been tested, namely (i) not using the latent space as input for each time step, (ii) not conditioning the timestamp LSTM to the current activity $\hat{a}_i$ and (iii) using a shared LSTM for both activities and timestamps. However, the first configuration yielded poor reconstructed control flows, whereas the second and third configurations resulted in poor timestamp reconstruction.

Categorical and numerical attributes reconstruction follows different paths, one for each attribute, composed of a sequence of two fully connected layers and ReLU activations.
For both activity and categorical attributes predictions, the model outputs a probability distribution for each possible value and the argmax operator is used to select the most likely one.

\subsection{Training}
\label{sec:approach-training}

We train the model end-to-end with backpropagation, optimizing the CVAE loss function \eqref{eq:cvae_loss}. In particular, the reconstruction loss of a reconstructed trace $\hat{x} \sim p_\theta(x|\mb{z},c)$ with respect to its ground-truth trace $x$ is the sum of the following loss components:\footnote{Using these reconstruction loss components corresponds to evaluating (i) the likelihood of every activity and categorical trace attribute against the respective categorical (softmax-normalized) distribution predicted by the decoder, and (ii) the likelihood of every timestamp and numerical trace attribute against a Gaussian distribution (representing a sort of additive noise) located at respective real-valued prediction returned by the decoder.}

\begin{itemize}

\item Binary Cross Entropy (BCE) loss of each trace activity 
\item Mean Squared Error (MSE) loss of each event interarrival time 
\item BCE loss of each categorical attribute of the trace
\item MSE loss of each numerical attribute of the trace
\end{itemize}
To prevent vanishing of the KL divergence loss we make use of a technique called \emph{KL cyclical annealing}, which varies $\beta$ following a linear cyclical schedule as training progresses \cite{fu2019cyclical}. When $\beta < 1$, the model is able to focus more on improving reconstruction. In our tests without cyclical annealing, the KL divergence loss decreased to nearly zero whereas the reconstruction loss did not improve much.

\subsection{Generation}
\label{sec:approach-generation}

After the model has been trained, new traces can be generated by randomly sampling a latent space vector $\mb{z}$ from the multivariate standard Gaussian distribution $\mathcal{N}(\Vec{0}, \mb{I})$, attaching a conditional variable to it and feeding the resulting vector to the decoder network. The conditional variable makes it possible to limit the generation of traces to the specified variable only.

The decoder activity LSTM recurrently generates activities for the trace until the EOT token gets generated or until a fixed maximum trace length is reached. For each activity, the LSTM outputs a probability distribution and the argmax operator is used to choose the most likely activity.

To reconstruct timestamps, the decoder makes use of three pieces of information, namely (1) an arbitrary start timestamp $\tau$, (2) the \emph{trace arrival time} $\hat{T}_1$ and (3) the \emph{event interarrival times} $\hat{t}_i$. The timestamp $\hat{T}_i$ of activity $\hat{a}_i$ is computed as follows:
$\hat{T}_i = \tau + \hat{T}_1 + \sum_{k=1}^{i} \hat{t}_k$.
The start timestamp $\tau$ can be chosen arbitrary, but it is usually set to the first timestamp of the entire log, or to the first or last timestamp of the test log. Also note that, given the way trace arrival times $T_i$ are processed, the model generates traces that follow the arrival time distribution of the training set.

\section{Evaluation}
\label{sec:Evaluation}

In this section, we present the evaluation methodology used to assess the outcomes of conditional generative models for trace generation.
We compare our method (denoted as \cvae) with the DL generative model from \cite{Camargo_lstm}.\footnote{We use the release \emph{First version refactory} in the repository cited in \cite{Camargo_lstm}.} 
We aim to address the following research questions:
\begin{enumerate}[label=\textbf{RQ\arabic*},leftmargin=3\parindent]
	\item \label{RQ1} Quality of Generated Traces: How is the quality of the generated traces, in terms of temporal and control-flow dimensions, compared to other state-of-the-art generative models?
	\item \label{RQ2} Variability vs. Compliance: What is the trade-off between the variability of the generated traces and their compliance with the original process, compared to other state-of-the-art generative models?
	\item \label{RQ3} Effectiveness of Conditional Control: How effective is the control provided by the conditional variable in guiding the generative model to produce specific types of traces?
\end{enumerate}
\ref{RQ1} aims to investigate the overall quality of the generation, focusing on three main aspects: the control-flow, the temporal distribution of events and the distribution of trace cycle times.
\ref{RQ2} aims at assessing the capability of the model to produce original traces different from the one of the training set, while keeping them meaningful with respect to
process constraints.
Finally, \ref{RQ3} aims at analysing the  effectiveness of the model's conditional mechanism and its ability to correctly reproduce the various types of traces defined by the conditional variable.

\begin{figure*}[!t]
    \centering
    \includegraphics[width=\linewidth]{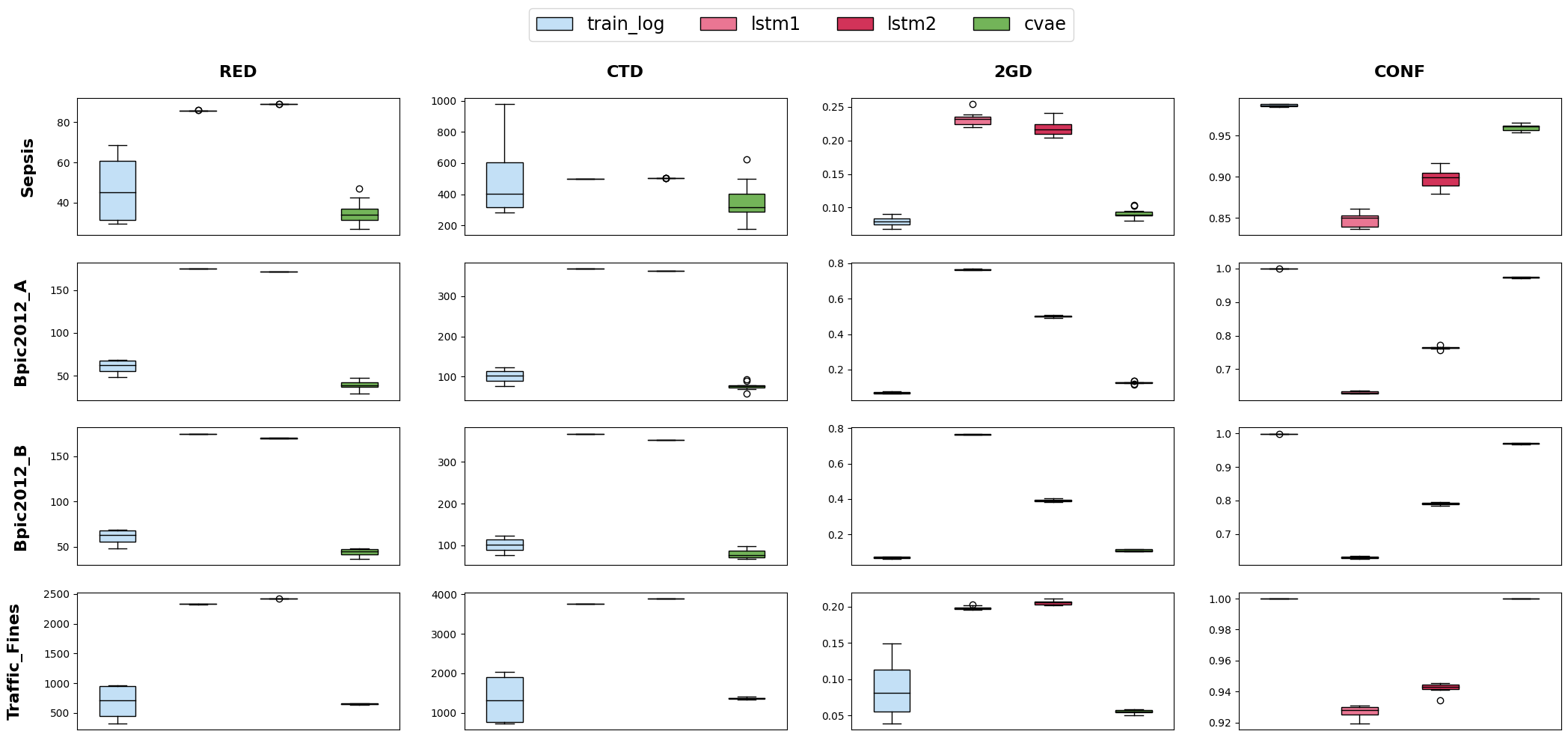}
    \caption{Conditioning-free generation (\textbf{RQ1}): results obtained for the proposed \cvae method, the competitors \lstmi and \lstmii and the ``optimistic'' baseline \baseline on the metrics, RED, CTD, 2GD (the lower the better), CONF (the higher the better).
    }
    \label{fig:evaluation}
\end{figure*}

\begin{table}[t!]
	\caption{Datasets Description}\label{tab:datasets}
	\centering
	\resizebox{.95\columnwidth}{!}{
		\begin{tabular}{lcccccc}
			\toprule
			\multirow{2}*{Dataset}  & \multirow{2}*{Trace \#} & \multirow{2}*{Variant \#}  &  \multirow{2}*{Activity \#}  & Avg. trace &	Avg. trace & Conditional  \\
			&  &  &  & length & cycle time & ratio  \\
			\midrule
			\sepsis  &  782 & 733 & 16 & 17 & 38.2d  &  14\%  \\
			\bpia  &  4685 & 3790 & 36 & 40 & 18.7d  &  17\%  \\
			\bpib &  4685 & 3790 & 36 & 40 & 18.7d  &  28\%  \\
			\fines &  129615  & 200 & 10 & 4 & 20.2wks  & 4\%   \\
			\bottomrule
		\end{tabular}}
\end{table}

\subsection{Datasets}
\label{sec:datasets}
Our evaluation considers four  examples based on three real-world event logs,
preprocessed to remove incomplete traces.
For each example, we define a binary conditioning on 
trace execution, that distinguishes between two relevant subprocesses, enabling us to generate only the desired traces by tuning the conditional variable.

\begin{itemize}
\item 
\textbf{Sepsis cases} \cite{sepsis_log} contains events of sepsis cases from a hospital. We filtered out incomplete traces with a missing ``Release'' activity. We consider the following conditional labelling: the patient has a relapse and returns to the emergency room within 28 days from discharge. We denote this example as \sepsis.
\item \textbf{BPIC2012} \cite{vandongen_2012} 
contains a loan application process.
We filtered out incomplete traces and those corresponding to ineligible applications, keeping those for which at least one offer was created. We consider two different conditional labellings for this log: (i) the reject of a loan offer by the applicant, indicated by the presence of the activity ``O\_DECLINED-COMPLETE'', denoted as \bpia; (ii) the creation of multiple loan offers by the bank, indicated by the presence of multiple ``O\_CREATED-COMPLETE''  events, denoted as \bpib.\footnote{The BPIC2012 log also contains information about the events' lifecycle transitions: ``SCHEDULE'', ``START'', and ``COMPLETE''. This information is concatenated with the activity label of each event in order to reproduce it.}
\item \textbf{Traffic fines} \cite{traffic_fines} is an event log of an information system managing road traffic fines. We filtered out incomplete traces, 
whose last activity is ``Send Fine''. The conditional labelling is defined by the presence of an appeal request by the offender to the judge or the prefecture. This corresponds to the presence of any of the following activities: ``Appeal to Judge'', ``Send Appeal to Prefecture'', ``Insert Data Appeal to Prefecture'', ``Notify Result Appeal to Offender'', ``Receive Result Appeal from Prefecture''.
This example is denoted as \fines.
\end{itemize}
In Table~\ref{tab:datasets}, we report details on the four datasets, highlighting their variety in terms of number of traces, activity labels, cycle times, and \emph{conditioning ratios}, i.e., the percentage of traces for which the conditional labelling is true.

\subsection{Methodology}
\label{sec:evaluation_method}

Each dataset is obtained by adding a trace attribute to every trace, containing the value of the conditional labelling, as defined in the previous section.
Each dataset is then split in training, validation and test set in chronological order.\footnote{In general, we apply a 70\%-10\%-20\% split, with the exception of \fines,  for which we take a 5\% test set that already contains 6480 traces, which is in the same order of magnitude of the other test logs.}
Our model is trained using the following hyperparameters: Embedding size (categorical attributes and activities) = 5, LSTM hidden size = 200, Latent space size = 10, Learning rate = $3 \times 10^{-4}$, Dropout = 5\%, Batch size = 256, Number of KL annealing cycles = 8.


During training, the validation set is used to activate an early stopping mechanism, which stops the training when the loss function, computed on the validation set, does not improve for 100 epochs, and outputs the model with the best loss. This only applies when evaluating the full loss function ($\beta=1$, i.e., KL annealing cycles are excluded by early stopping). 

The trained model is used to generate 10 different logs, which are used for the assessment. Each generated log contains the same number of traces as the corresponding test set.
The generation is guided by setting the conditional variable to reproduce the same conditional ratio found in the training log.

\subsection{Metrics}
\label{sec:metrics}

We present the framework for the evaluation of the generative model.
To assess the quality of the generated traces, we adopt a subset of the metrics introduced in \cite{Dumas_framework} within the domain of simulation models.
These metrics quantify the ``dissimilarity'' between the generated traces and those in the test set by computing the Earth Mover's Distance (EMD) of their respective distributions along different temporal and control flow dimensions:
\begin{itemize}
\item Event time distribution within the trace, in the case of metric \emph{Relative Event Distribution} (RED);\footnote{We adopt the RED metric instead of the Absolute Event Distribution (AED) metric from \cite{Dumas_framework} because we want to assess the quality of individual generated traces, focusing on the temporal distribution of events within the trace, rather than on the overall log temporal horizon.}
\item Cycle times, in the case of metric \emph{Cycle Time Distribution} (CTD);
\item Event class N-grams, in the case of metric \emph{$N$-Gram Distance} (NGD), which we specifically computed by setting $N=2$ (2GD), using the directly-follow graphs of the generated and test traces. 
\end{itemize}

Inspired by the work in couterfactual generation in PM \cite{buliga2023counterfactuals}, we consider a further metrics measuring the compliance of the generated traces with the process implicit constraints:
\begin{itemize}
\item \emph{Conformance score} (CONF) measures the average number of \textsc{declare} constraints \cite{declare} satisfied by the generated traces. The \textsc{declare} constraints are mined from the event log with a support of 90\%.%
\footnote{To ensure fairness, only generated variants that do not already appear in the training log are used for this analysis.}
\end{itemize}

To quantify the \emph{variability} of the generated traces we focus on the control-flow and compute the number of new variants generated with respect to those in the training and test log.

Finally, to assess the conditional generation we 
compute the \emph{actual conditional ratio} of the generated traces. To do so we recompute a-posteriori the conditional labelling, as defined in Sect.~\ref{sec:datasets}, and compare the conditional rate with the ones of the training and test log.

\subsection{Benchmarks}
\label{sec:benchmarks}

As mentioned at the beginning of this section we compare our method with the LSTM model introduced in \cite{Camargo_lstm}, which has been identified as the state-of-the-art deep generative model for PM in \cite{CamargoDumas_survey}. 
Since this method is not aware of the conditional variable, we leverage it in the following two ways: 
\begin{itemize}
\item we train it on the full training set and use it to unconditionally generate traces. We denote this method as 
\lstmi;
\item we train two separate models on the two subsets of the training data identified by the value of the binary conditional variable. We then generate traces using both models to reproduce the conditional ratio of the training set. We denote this method as \lstmii.
\end{itemize}

To provide an additional point of comparison, we also consider the original log itself as a baseline reference for the metrics RED, CTD, and 2GD. To achieve this, we split the training plus the validation set into four parts based on the chronological order. Each of these components contains the same number of traces as the test set, thus enabling a direct comparison when computing the aforementioned metrics. We denote this baseline as \baseline.

\section{Results}
\label{sec:Results}
In this section, we present the obtained results.\footnote{The implementation to reproduce the experiments, the datasets, and the additional results --- including the temporal distributions of each activity and trace attribute value as well as other evaluation metrics --- can be accessed at the following link: \url{https://github.com/rgraziosi-fbk/cvae-process-mining}.}
We start by showcasing an example of conditional generation with \cvae in Table~\ref{tab:generation}, where we list the control-flow of two pairs of traces generated for the experiment \fines. Each pair of traces is  generated by the same value of the latent variable $\mb{z}$ but with different values of the conditional variable $c$.
These two examples can be interpreted as ``what-if'' scenarios. When $c=\text{F}$, in both cases the fine is paid by the offender within a short time. When $c=\text{T}$, the offender appeals in the two traces to the Judge and to the Prefecture, respectively. 
This causes delays, and, as a consequence, a penalty is added six months after from the issuance of the fine, and this is known to be a process constraint \cite{traffic_fines_report} --- when answering \ref{RQ3} we will show that the robustness under process constraints is not an accident of this example, but is a general feature of our model.
The first example ends with the \emph{Payment} activity, indicating that the appeal has been denied. In contrast, the second example shows that the appeal to the prefecture has been accepted, and no payment is due.

 \begin{table*}[t!]
\caption{Example of generations}\label{tab:generation}
    \centering
    \resizebox{\textwidth}{!}{
        \begin{tabular}{c|l}
            \toprule
            $c$ & trace \\
            \midrule
            F & $\langle$ (Create Fine, 29/12/99) , (Payment, 9/1/00) $\rangle$ \\
            T & $\langle$ (Create Fine, 1/1/00), (Send Fine, 15/4/00), (Insert Fine Notification, 27/4/00), (Appeal to Judge, 27/5/00), (Add penalty, 29/6/00), (Payment, 23/9/01) $\rangle$ \\
            \midrule
            F & $\langle$ (Create Fine, 2/1/00), (Send Fine, 12/3/00), (Payment, 12/4/00) $\rangle$ \\
            T & $\langle$ (Create Fine, 1/1/00), (Send Fine, 15/4/00), (Insert Fine Notification, 29/4/00), (Insert Date Appeal to Prefecture, 6/6/00), (Add penalty, 28/6/00), (Send Appeal to Prefecture, 21/7/00) $\rangle$ \\
            \bottomrule
        \end{tabular}
    }
\end{table*}

We start the analysis of the results by answering \ref{RQ1}. In Figure~\ref{fig:evaluation} we plot the values of the temporal (RED, CTD) and control-flow (2GD) metrics computed for each of the four experiments described in Section~\ref{sec:datasets}.
We can observe that \cvae consistently outperforms the competing \lstm models in all three metrics: relative event distribution (RED), trace cycle time (CTD) and 2-grams (2GD).
It is also worth noting that the \cvae boxplots often overlap with the \baseline ones, indicating that it correctly reproduces the distributions of the training set for these three metrics.
In summary, the quality of the traces generated by the \cvae appears to be much higher than those of the \lstm models on both the control-flow and temporal aspects.
Finally, we observe that across different logs and metrics, the boxplots of all three generative models are significantly narrower than the ones of the baseline.
Compared to a pure data sampling approach (as in the baseline), these  models appear more stable in producing representative trace samples. However, all three methods likely fail to capture the full variability of the original logs.

To address \ref{RQ2}, we report in Table~\ref{tab:variants} the average number of variants generated by the models that already appear in the training or test sets. We observe that, compared to the \lstmi and \lstmii models, which almost always generate new variants, 
depending on the dataset, from 15\% to 77\% (from 1\% to 42\%) of the variants generated by the \cvae model already appear in the training log (test log).
Looking at the CONF values reported in Figure~\ref{fig:evaluation}, we see that \cvae achieves a significantly higher conformance, ranging from 5\% to more than 30\% higher than the \lstm models and very close to the maximal ``self''-conformance value (reached by the \baseline baseline). 
This indicates that almost all the variability generated by the \cvae model complies with the process implicit constraints, while the \lstm models introduce a noticeable degree of non-conformity.
This difference is likely due to the different ``hallucination'' mechanisms used by the two architectures to generate diverse outputs. Specifically, the sampling in the latent space used by the \cvae is more robust compared to the random choice selection of the LSTM network's output used by the \lstm models. The latter method has a relatively small but appreciable probability of sampling an incongruous next activity during the trace generation.
\begin{table}[t!]
	\caption{Variant analysis}\label{tab:variants}
	\centering
	\resizebox{.9\columnwidth}{!}{
		\begin{tabular}{l|l|ccc}
			\toprule
			\multirow{2}*{Model} & \multirow{2}*{Dataset}  & Variant \# & Variant \#  &  Variant \#   \\
		&	& (total)  & (in training) & (in test) \\
			\midrule
                \multirow{4}*{\cvae} & \sepsis & 152.7 & 52.8 & 2 \\
                 & \bpia & 880.4 & 125.1 & 43.8 \\
                 & \bpib & 882.2 & 153 & 48.9 \\
                 & \fines & 60.7 & 46.6 & 25.6 \\
                \midrule
                \multirow{4}*{\lstmi} & \sepsis & 157 & 0.1 & 0.1 \\
                  & \bpia  & 937 & 0 & 0 \\
                  & \bpib  & 937 & 0 & 0\\
                  & \fines & 282.7 & 16.1 & 9.8 \\
                \midrule
               \multirow{4}*{\lstmii} & \sepsis & 157 & 0.8 &  0.0\\
                  & \bpia & 936.6 & 1 & 0.6 \\
                  & \bpib & 936.6 & 0.3 & 0.3 \\
                  & \fines & 218.6 & 42 & 26.5 \\
			\bottomrule
		\end{tabular}}
\end{table}

Finally, we address \ref{RQ3}.
Table~\ref{tab:conditional_ratio} reports the original conditional ratios for the training and test logs, as well as those computed a-posteriori for generated logs. 
This analysis, using the definition from Sect.~\ref{sec:datasets}, determines the conditional labels of the generated traces. It helps assess whether the generative models accurately reproduce the labels they were constrained by during inference and identifies any discrepancies in label reproduction.
We expect that the generated logs reproduce the ratio observed in the training set.
We observe that \cvae consistently reproduces a conditional ratio that differs from the training ratio of at most 2\%.
This is not the case for none of the two \lstm models. 
While \lstmii generally performs well, it fails to reproduce the correct ratio specifically in the \sepsis log, despite being trained separately on the two conditioned subsets of the log.
Notably, the \sepsis log is the only example that includes temporal conditioning.
The robustness of the conditional generation of the \cvae model is further confirmed by the computation of the RED, CTD, 2GD e CONF 
metrics separately on the two log subsets.\footnote{The results of the analysis are available at the reproducibility link.}
\begin{table}[t!]
\caption{Conditional ratio}\label{tab:conditional_ratio}
    \centering
        \begin{tabular}{lccccc}
            \toprule
            dataset & training & test & \cvae & \lstmi & \lstmii  \\
            \midrule
            \sepsis & 14.8\% & 11.5\% & 15\% & 28.6\% & 38.3 \% \\
            \bpia & 16.5\% & 19.2\% & 15.3\% & 14.4\% & 15.1\% \\
            \bpib & 28.2\% & 26\% & 27\% & 31.5\% & 27.3\% \\
            \fines & 3.2\% & 3.6\% & 4\% & 3\% & 4\% \\
            \bottomrule
        \end{tabular}
\end{table}

\section{Conclusions}
\label{sec:Conclusions}

In this paper, we introduced a conditional variational auto encoder for trace generation.
We showed that our method outperforms current state-of-the-art generative models for trace generation in terms of the quality of the traces' control-flow, cycle time and event temporal distribution, as well as of their compliance with process implicit constraints.
Moreover, we showed that it is possible to robustly control the generation process by setting the conditional variable, so as to generate only the traces of interest or to simulate ``what-if'' scenarios.
In the future, we plan to extend the log generation by also taking into account resources and to improve the reproduction of the traces temporal distribution.



\bibliographystyle{IEEEtran}
\bibliography{IEEEabrv,biblio}

\end{document}